\documentclass[structabstract,printer]{aa}

\usepackage{graphicx}
\usepackage{txfonts}
\usepackage{epsf}
\usepackage{epsfig}

\def \etal{et~al.\/}
\def\lesssim{\mathrel{\hbox{\rlap{\hbox{\lower4pt\hbox{$\sim$}}}\hbox{$<$}}}}
\def\gtrsim{\mathrel{\hbox{\rlap{\hbox{\lower4pt\hbox{$\sim$}}}\hbox{$>$}}}}

\begin{document}

\title{Scattering polarization due to light source anisotropy}

\subtitle{II. Envelope of arbitrary shape}

\author{R.~Ignace\inst{1}, M.~B.~Al-Malki\inst{2,3},
J.~F.~L.~Simmons\inst{2} J.~C.~Brown\inst{2}, D.~Clarke\inst{2}, \and
J.~C.~Carson\inst{1}}

\institute{
    Department of Physics \& Astronomy, East Tennessee
    State University, Johnson City, TN, USA
    \and
    Kelvin Bldg, Department of Physics and Astronomy, University of Glasgow,
        Glasgow, G12 8QQ, Scotland UK
    \and
    Now at P.O. Box 87946, Riyadh 11652 Saudi Arabia}

\offprints{ignace@etsu.edu}

\date{Received <date>; Accepted <date>}

\authorrunning{Ignace \etal}

\titlerunning{Polarization from arbitrary envelopes}

\maketitle

\abstract{}
{We consider the polarization arising from scattering in an envelope
illuminated by a central anisotropic source.  This work extends the
theory introduced in a previous paper (Al-Malki \etal\ 1999) in
which scattering polarization from a spherically symmetric envelope
illuminated by an anisotropic point source was
considered. Here we generalize to account for the more realistic
expectation of a non-spherical envelope shape.}
{Spherical harmonics
are used to describe both the light source anisotropy and the
envelope density distribution functions of the scattering
particles.  This framework demonstrates how the
net resultant polarization arises from a superposition of three
basic ``shape'' functions:  the distribution of source illumination,
the distribution of envelope scatterers, and the phase function for
dipole scattering.}  
{Specific expressions for the Stokes parameters
and scattered flux are derived for the case of an
ellipsoidal light source inside an ellipsoidal envelope, with
principal axes that are generally not aligned.
Two illustrative examples are considered:  (a) axisymmetric
mass loss from a rapidly rotating star, such as may apply to some
Luminous Blue Variables, and (b) a Roche-lobe filling star in
a binary system with a circumstellar envelope.}
{As a general conclusion, the combination of source anisotropy with
distorted scattering envelopes leads to more complex polarimetric
behavior such that the source characteristics should be carefully
considered when interpreting polarimetric data. }

      \keywords{
           Polarization --
           Stars: circumstellar --
           Stars: binaries --
           Stars: rotation --
           Stars: winds
                }

\section{Introduction}

For many stars linear polarization is produced mainly from scattering
of starlight by circumstellar matter (Kruszewski \etal\ 1968; Serkowski
1970; Dyck \etal\ 1971; Shawl 1975).  This polarization can be used
as a diagnostic of the geometry of the circumstellar envelope and of
the light source(s) (e.g., Shakhovskoi 1965; Serkowski 1970; Brown \&
McLean 1977; Brown \etal\ 1978; Rudy \& Kemp 1978; Simmons 1982, 1983;
Friend \& Cassinelli 1986; Clarke \& McGale 1986, 1987).  In many models
for circumstellar scattering polarization, the illuminating sources are
treated as isotropic point sources (e.g., Brown \& McLean 1977;
Brown \etal\ 1978; Rudy \& Kemp 1978; Shawl 1975; Simmons 1982).
The effect of the finite size of the star as a light source has been
studied (Cassinelli, Nordsieck, \& Murison 1987), including the effects
of limb darkening (Brown \etal\ 1989) and stellar occultation (Brown \&
Fox 1989; Fox \& Brown 1991; Fox 1991).  Variable polarization is also
revealing about a system, as evidenced in a recent study by Elias,
Koch, \& Pfeiffer (2008).  With recent emphasis on clumped wind flows
of hot stars (e.g., Hamann, Feldmeier, \& Oskinova 2008), models have
been developed to interpret variable polarization from hot star winds
(Richardson, Brown, \& Simmons 1996; Brown, Ignace, \& Cassinelli 2000;
Li \etal\ 2000; Davies \etal\ 2007). 

Relatively little has been done to explore the anisotropy
of the source illumination and its consequences for interpreting
polarimetric observations.  There are some exceptions, such as a study of
gravity-darkening effects for the polarization from Be~stars by Bjorkman
\& Bjorkman (1994), the influence of star spots for polarizations
from pre-main sequence stars by Vink \etal\ (2005), and more recently
calculations of source anisotropy for interpreting polarimetric behavior
observed in the post-red supergiant star HD~179821 by Patel \etal\
(2008).  In a previous paper (Al-Malki \etal\ 1999; hereafter Paper I),
we modeled the polarization arising from Thomson and Rayleigh scattering
explicitly for an arbitrary anisotropic (point) light source.  As a proof
of concept, the circumstellar envelope was taken as spherically symmetric
to explore the polarization signals that arise solely from the properties
of the source.  An upper limit to the polarization of about 20\% was
derived in the limiting case of a disk-like star when viewed edge-on.

This paper generalizes the results of Paper~I by allowing both the point
light source and the particle density distribution to be functions
of arbitrary angular form.  As in Paper~I, we restrict ourselves to
optically thin envelopes and single Thomson or Rayleigh scattering
mechanisms. Although the effects of a finite-sized illuminator will be
important for obtaining quantitatively a more accurate polarization
amplitude (e.g., Ignace, Henson, \& Carson 2008), the point source
approximation is certainly adequate for exploring polarization trends
of generalized source and envelope geometries.  In Sect.~2, we develop
the theory required to evaluate the polarization, based on Paper I and
the discussion of Simmons (1982). Applications to specific cases of
astrophysical interest are explored in Sect.~3, with emphasis given to a
treatment of the star and the circumstellar envelope as ellipsoidal. A
discussion of these results and the need for ongoing modeling is given
in Sect.~4.

\begin{figure*}[t]
\centering{\epsfig{file=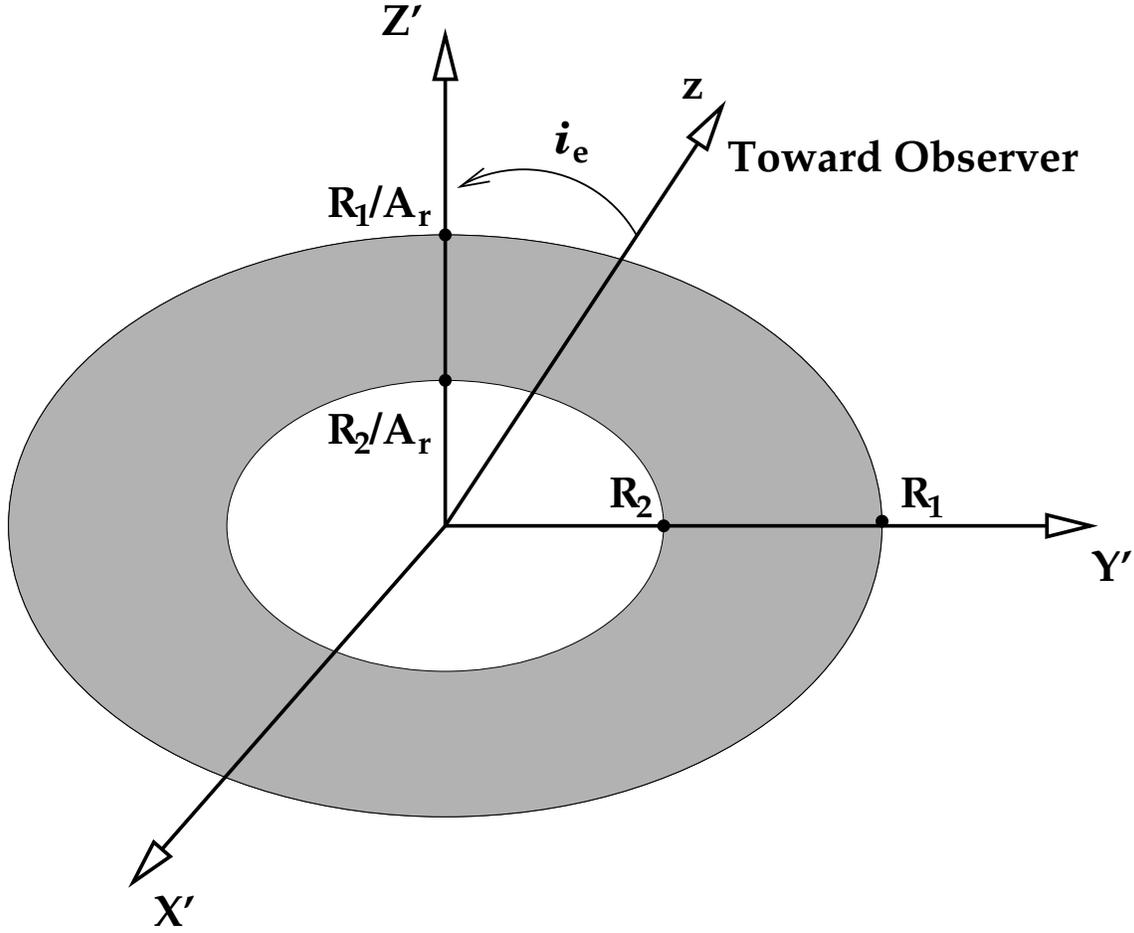,angle= -90,width=15cm}}
\caption{Illustration of the ellipsoidal envelope used in our models.
The scattering envelope is shown in cross-section as shaded, with
an equatorial width ranging from an inner radius of $R_2$ to an
outer one of $R_1$.  Along the pole, the inner and outer boundaries
are scaled by the factor $A_{\rm r}$.  Although the theory allows
for fully triaxial ellipsoids, our specific applications are for
axisymmetric ellipsoids (ranging from oblate to prolate).
\label{fig1}}
\end{figure*}

\section{General expression for scattered flux and polarization}

As in Paper I, we neglect the effects of finite star depolarization
and stellar occultation.  Consequently scattering particles in the
circumstellar envelope ``see'' a point star.  However, we model the
anisotropy of the illumination with a ``shape'' function to represent
directional flux.

Under these approximations,
our goal is to derive the net polarization from dipole scattering
for an unresolved system that allows for an arbitrary circumstellar
envelope geometry and source geometry.  Our approach is to represent
the different geometrical factors in terms of expansions in
spherical harmonics.  As a result, it is important to be clear about
the different coordinate systems being employed. 

Following the notation of Paper~I, a description of the problem
requires three coordinate systems, each centered on the star:  one
for the observer, one for the envelope, and one for the illumination
pattern of the star itself.  In general, none of these systems
are collinear.
The systems are defined as follows:

\begin{itemize}

\item[i)] For the observer reference frame we define Cartesian
coordinates $(x,y,z)$ centered at the star, with spherical
coordinates $(r,\theta,\phi)$, the line-of-sight being the
$Oz-$axis. Thus $\hat{z}$ is a unit vector in the direction of the
observer from the star and $x,y$ are observer coordinates in the
plane of the sky. For a scattering point in direction $\hat{r}$, the
scattering angle is given by $\cos \theta = \hat{z}\cdot\hat{r}$, as
in Paper~I.  The angle $\phi=\tan^{-1} y/x$ is the
observer's polarization angle (orientation) for any
scattering point.

\item[ii)] The star's frame is $(X,Y,Z)$ with spherical coordinates
$(r,\vartheta,\varphi)$, where $OZ$ is a convenient stellar axis
(such as a rotation axis) that lies in the $x-z$
plane of the observer's frame. The star system is rotated relative
to the observer coordinates $(x,y,z)$ through standard Euler angles
$(\alpha, \beta, \gamma)$.

\item[iii)] The envelope frame $(X',Y',Z')$ with spherical coordinates
$(r,\vartheta',\varphi')$, centered on the star, where $OZ'$ is a
convenient axis for the envelope.

\end{itemize}

In general the scatterer density is $n(r,\vartheta',\varphi')$, and
the flux of radiation from the star is $F(r,\vartheta,\varphi)$.
The stellar illumination is taken to be unpolarized.  In fact,
stellar atmospheres may show some low level of intrinsic polarization
(Collins 1970);
however, the polarization from the circumstellar scattering will be
dominated by the Stokes-I component from the
source.  Following Eq.~(1) from Paper~I
(see also Simmons 1982), the scattered flux and Stokes parameters
$(F_{\rm sc}, Q, U)$ of the scattered radiation at the Earth (distance
$D$) are given in the form:

\begin{eqnarray}
\left. \begin{array}{c} F_{\rm sc} \\ Z^* \end{array}\right\}  & = &
    \frac{1}{2k^2\,D^2}\,\int\int\int\,n(r,\theta',\phi')\,
    F(r,\vartheta,\varphi)\,r^2\times  \nonumber \\
 & & \left\{ \begin{array}{l} (i_1+i_2) \\ (i_1-i_2)\,\exp(-2i\phi)
    \end{array} \right. \, dr\,\sin\theta\,d\theta\,d\phi,
    \label{eq:stokes}
\end{eqnarray}

\noindent with the same definitions as in Paper~I, with $Z^*=Q-iU$, for
$i=\sqrt{-1}$, $k=2\pi/\lambda$ the wave number, and $i_1$ and $i_2$
the scattering functions as defined by van de Hulst (1957).  The
circular polarization $V$ is assumed zero.  For Thomson (free electrons) or
Rayleigh scattering, we have

\begin{equation}
i_1 \pm i_2 = \frac{3k^2}{8\pi}\,\sigma\,(1\pm\cos^2\theta),
\end{equation}

\noindent where the value of the cross section factor $\sigma$ is
chosen according to whether Thomson scattering ($\sigma$ independent
of $k$) or Rayleigh scattering ($\sigma \propto k^4$) is considered.

Provided that $F$ varies smoothly, it may be expressed in terms of
spherical harmonics in the observer frame (see
Paper~I):

\begin{equation}
F(r,\theta,\phi) = \sum_{l=0}^{\infty}\,\sum_{{\rm m}=-l}^{{\rm m}=l}\,
    F_{l{\rm m}}(r)\,\sum_{{\rm n}=-l}^{{\rm n}=l}\,R^{l}_{\rm nm}
    (\alpha,\beta,\gamma)\,Y_{l{\rm n}}(\theta,\phi),
    \label{eq:Fexpand}
\end{equation}

\noindent where

\begin{equation}
F_{l{\rm m}} (r) = \int^1_{-1}\,\int_0^{2\pi}\, F(r,\theta,\phi)\,
        Y^*_{l{\rm m}}(\theta,\phi) \,d(\cos\theta)\,d\phi,
        \label{eq:Flm}
\end{equation}

\noindent and $R^l_{nm}$ represents elements of a rotational matrix
(see Paper~I).

Similarly, the density distribution of scatterers can also be expanded in
terms of spherical harmonics, which becomes

\begin{equation}
n(r,\theta,\phi) = \sum_{l'=0}^{\infty}\,\sum_{{\rm m'}=-l'}^{{\rm m'}=l'}\,
     n_{l'{\rm m'}}(r)\, Y_{l'{\rm m'}}(\theta,\phi),
    \label{eq:nexpand}
\end{equation}

\noindent where

\begin{equation}
n_{l'{\rm m'}}(r) = \int^1_{-1}\,\int_0^{2\pi}\, n(r,\theta',\phi')\,
        Y^*_{l'{\rm m'}}(\theta',\phi') \,d(\cos\theta')\,d\phi'.
        \label{eq:nlm}
\end{equation}

\noindent The $Y_{l{\rm n}}(\theta,\phi)$ factors in
Eqs.~(\ref{eq:Fexpand}) and (\ref{eq:nexpand}) are as defined in
Paper~I (c.f., Jackson 1975).  For the product of two spherical
harmonics, we can express the multipoles of $n(r,\theta,\phi)$ and
$F(r,\theta,\phi)$ as:

\begin{eqnarray}
n(r,\theta,\phi)\,F(r,\theta,\phi) & = & \sum_{ll'\,{\rm mm'}} \,
    n_{l'{\rm m'}}(r)\,F_{l{\rm m}}(r)\,\times  \nonumber \\
 & & Y_{l'{\rm m'}}(\theta,\phi)\,Y_{l{\rm m}}(\theta,\phi),
\end{eqnarray}

\noindent or as

\begin{equation}
n(r,\theta,\phi)\,F(r,\theta,\phi) = \sum_{l{\rm m}n} \, R^l_{\rm nm}
    \, \sum_{l'{\rm m'}} \, C^{\rm LM}_{ll'\,{\rm nm'}}\,
    Y_{\rm LM}(\theta,\phi).
    \label{eq:multipoles}
\end{equation}

\noindent In this latter expression, the set of $C^{\rm LM}_{ll'\,{\rm nm'}}$ 
values are
Clebsh-Gordon coefficients, arising from the products of two spherical
harmonics.  Only terms satisfying the following two conditions contribute
to the sum in Eq.~(\ref{eq:multipoles}):

\begin{equation}
{\rm n}+{\rm m'} = M,   \label{eq:cond1}
\end{equation}

\noindent and,

\begin{equation}
|l-l'| \le l \le l+l'.  \label{eq:cond2}
\end{equation}

\noindent The coefficients are given by the relation,

\begin{eqnarray}
C^{\rm LM}_{ll'\,{\rm nm'}} & = & (-1)^{\rm M}\,\sqrt{\frac{(2l+1)(2l'+1)
    (2L+1)}{4\pi}} \times \nonumber \\
 & & \, \left( \begin{array}{ccc} l & l' & L \\ 0 & 0
    & 0 \end{array} \right)
 \,\left( \begin{array}{ccc} l & l' & L \\
    n & m' & M\end{array} \right)
    \label{eq:clebsh}
\end{eqnarray}

\noindent (c.f., Messiah 1962 and the Appendix).

As in Paper~I,
the scattering factors can be expressed as

\begin{equation}
1+\cos^2\theta = \frac{4}{3}\,\left[ \sqrt{4\pi}\,Y_{00} +
    \sqrt{\frac{\pi}{5}}\,Y_{20}(\theta) \right],
\end{equation}

\noindent and

\begin{equation}
\sin^2\theta \, \exp(-2i\phi) = 4\,\sqrt{\frac{2\pi}{15}}\,Y_{22}^* (\theta,\phi),
\end{equation}

\noindent so that, upon substitution in the integrals contained in
Eq.~(\ref{eq:stokes}), together with the other expressions above, and using
the properties of spherical harmonics, $F_{\rm sc}$ and $Z^*$ can
be rewritten as:

\begin{eqnarray}
F_{\rm sc} & = & \frac{\sigma}{4\pi\,D^2}\,\sum_{l {\rm m n}} \, R^l_{\rm nm}
    (\alpha,\beta,\gamma)\times  \nonumber \\
    & & \left[ \sqrt{4\pi}\,C_{ll'nm'}^{00} + \sqrt{\frac{\pi}{5}}\,
    C_{ll'nm'}^{20}\right]\,S_{ll'mm'},
    \label{eq:Fsc}
\end{eqnarray}

\noindent and

\begin{equation}
Z^* = \frac{3\sigma}{4\pi\,D^2}\,\sqrt{\frac{2\pi}{15}}\sum_{l {\rm m n}}
    \, R^l_{\rm nm}(\alpha,\beta,\gamma) \,\sum_{l' {\rm m'}}
    C_{ll'nm'}^{22}\,S_{ll'mm'},
    \label{eq:Zstar}
\end{equation}

\noindent where

\begin{equation}
S_{ll'mm'} = \int^\infty_0 \, F_{l{\rm m}} (r) \, n_{l'{\rm m'}}(r) \,r^2\,dr.
    \label{eq:Sfunc}
\end{equation}

%
%
%

\noindent Thus, the 
scattered flux and the Stokes parameters can be expressed as sums
of increasing order of multipole contribution.  This helps to
separate the effects of anisotropy in the flux $F$ and in the
density distribution $n$, and to study how they can lead
to the production of polarization.

If the functions are smooth, the summations will converge rapidly,
so the first few terms ($l$ and $l' \le 2$) should provide reasonable
approximations for the behavior of the polarization.  Due to the
conditions of Eqs.~(\ref{eq:cond1}) and (\ref{eq:cond2}), the summation
over $l,l'$, m, and m' in Eqs.~(\ref{eq:Fsc}) and (\ref{eq:Zstar})
will be limited.  For example, if
the stellar flux and the circumstellar density
distribution of scattering particles is symmetric about stellar and
envelope polar axes respectively , the values of $S_{ll'{\rm mm'}}$
will be zero for odd values of $l$ and $l'$.

For a spherical envelope with $n(r,\theta,\phi)=n(r)$ and an
anisotropic light source, Eqs.~(\ref{eq:Fsc}) and (\ref{eq:Zstar})
reduce to the forms discussed in Paper~I. On the other hand, for an
isotropic light source within an envelope of arbitrary shape, the
expressions of Brown \& McLean (1977) and Simmons (1982) are
recovered.  

\begin{figure*}[t]
\centering{\epsfig{file=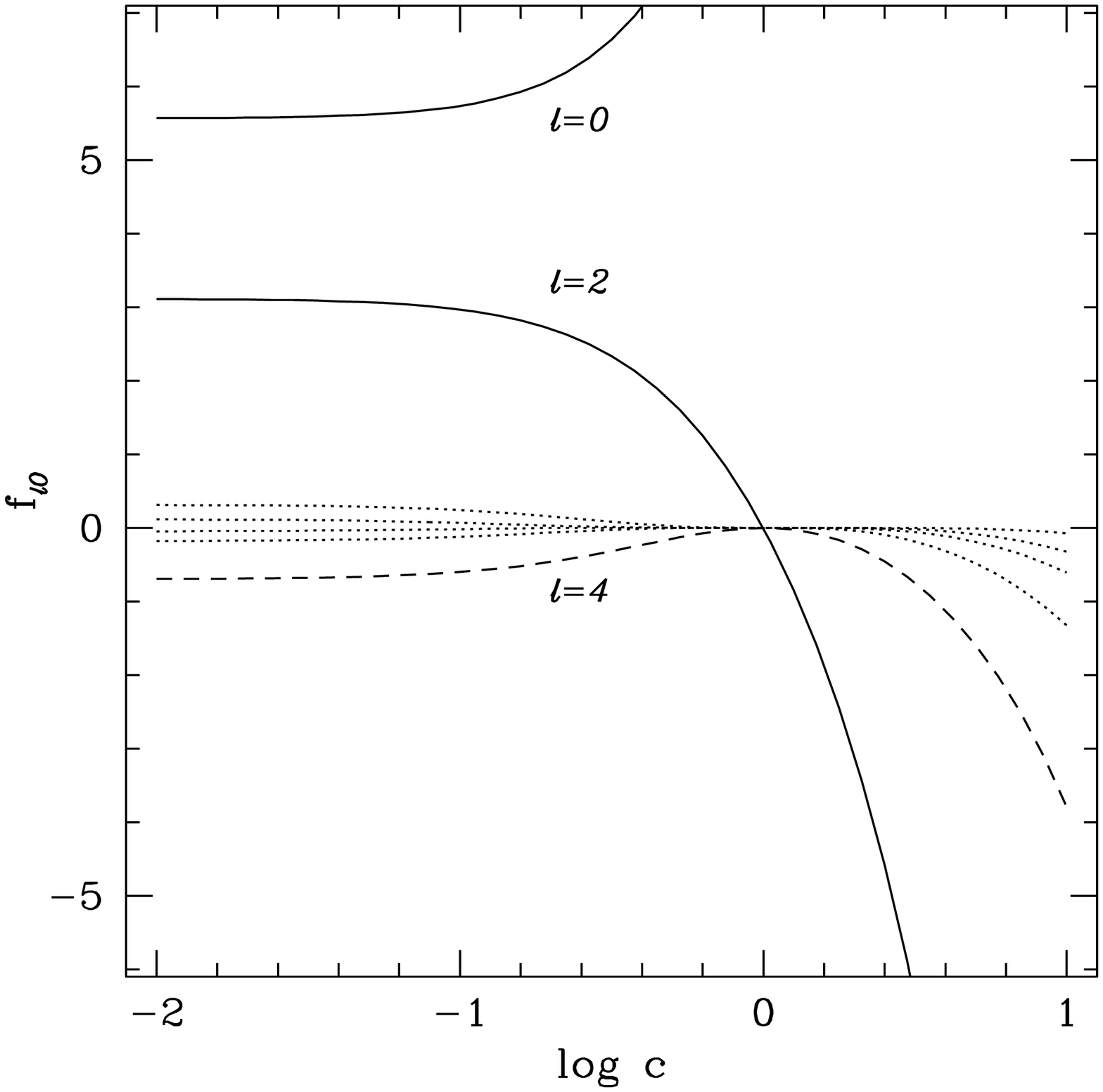,height=8cm} \epsfig{file=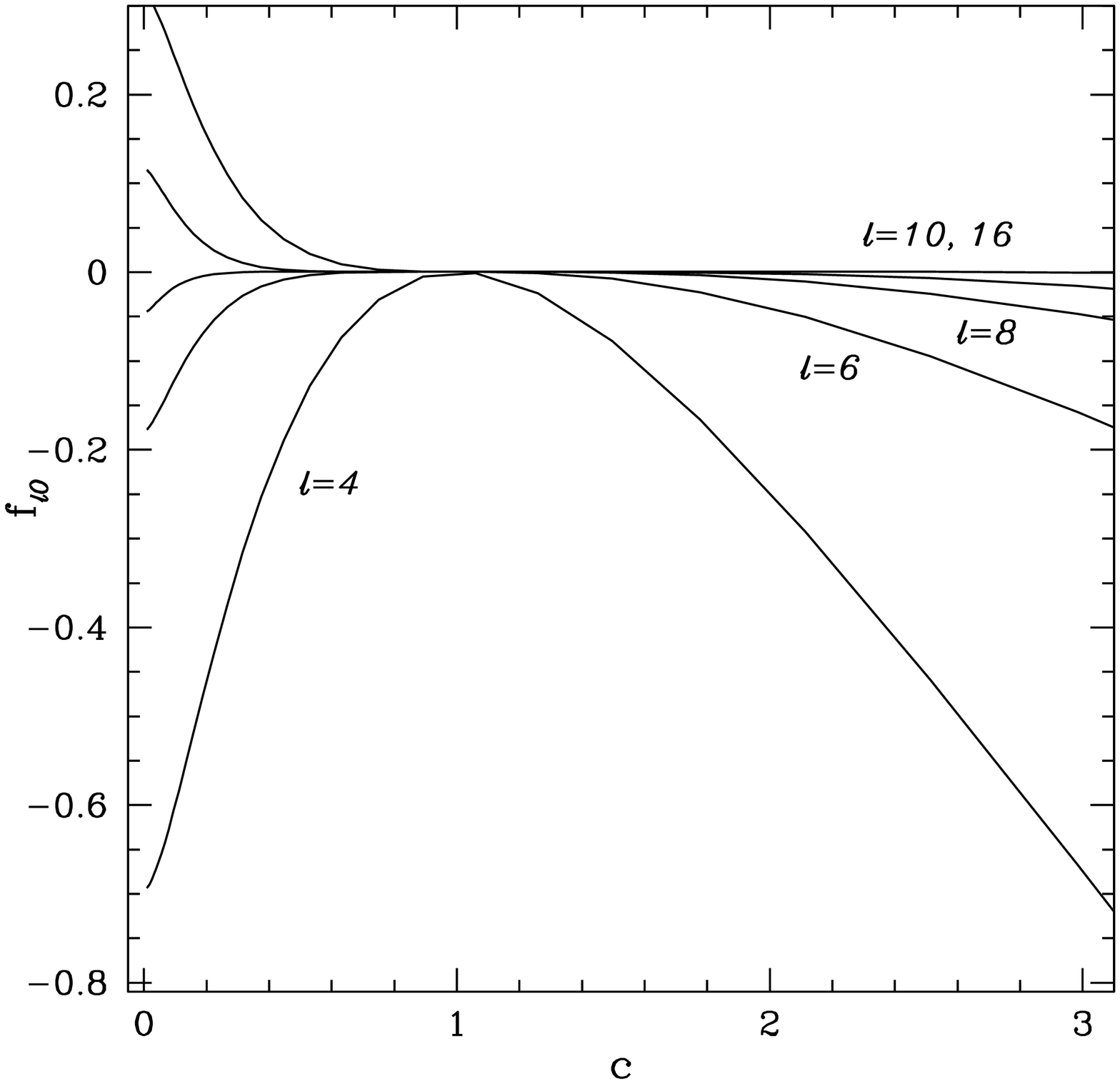,height=8cm}}
\caption{Plots of $f_{l0}$ for the source flux plotted against
the parameter $c$ for the anisotropy of the source.  Low $c$ implies
greater flux along the stellar poles, whereas large $c$ is the opposite
case.  The left panel shows curves for $l=0$ and $l=2$ as solid,
$l=4$ as dashed, and higher $l$ values as dotted (unlabeled) against
$\log c$.  The right panel shows a blow up around the region of $c=1$
and explicitly labels the higher order curves.  Certainly for $c<1$,
$l=0$ and 2 are most important.
\label{fig2}}
\end{figure*}

\section{Ellipsoidal light source and ellipsoidal circumstellar envelope}

As an illustration of the preceding general formulation, we will
take the case of a star of uniform surface brightness, as discussed
in Paper~I,  but now embedded within an ellipsoidally shaped
circumstellar envelope.  For an isotropic surface
intensity $I_*$, the flux $F(r,\vartheta, \varphi)$ can be expressed
(as seen from a distant point) in terms of the projected area
$A_{\rm p}(\vartheta, \varphi)$ of the star as seen from direction
$(\vartheta, \varphi)$. In other words we treat the star as an
arbitrarily small unpolarized illuminator 
but describe its anisotropy in terms of a distorted finite shape.

Under this approximation, the flux is given by

\begin{equation}
F(r,\vartheta, \varphi)\approx I_*\,\Delta \Omega = I_*\,
    \frac{A_{\rm p}(\vartheta, \varphi)}{r^2},
    \label{eq:ptsource}
\end{equation}

\noindent where $I_*$ is the isotropic intensity of the stellar surface
and $\Delta \Omega$ is the solid angle subtended by the scattering
element located a distance $r$ from the star.  As a concrete example,
we follow Paper~I (see their Fig.~2) by introducing star axes $(a,b,c)$
to be along $(X,Y,Z)$.  From Paper~I, the areal function is

\begin{equation}
A_{\rm p} = \pi \left| \sqrt{(bc\lambda)^2+(ac\mu)^2+(ab\nu)^2}\right| ,
    \label{eq:areal}
\end{equation}

\noindent where $(\lambda,\mu, \nu) = (\cos \varphi \,\sin\vartheta,
\sin\varphi\,\sin\vartheta, \cos \vartheta)$ are the $(X,Y,Z)$ direction
cosines for the direction $(\vartheta,\varphi)$.  So we obtain

\begin{equation}
F_{l{\rm m}} (r) = \frac{I_*}{r^2}\,\int_0^\pi\,\int_0^{2\pi}\,
    A_{\rm p}(\vartheta, \varphi)\,Y^*_{l{\rm m}}(\vartheta,\varphi)
    \,\sin \vartheta\,d\vartheta\,d\varphi.
    \label{eq:Fellipse}
\end{equation}

For the purpose of illustrative calculations, we now adopt simplifications
for the description of the illumination pattern and envelope density
profile.  Following Simmons (1982), we model the envelope as an
axisymmetric ellipsoidal shell (oblate or prolate) of arbitrary thickness
and constant density. The shell axis of rotational symmetry $OZ'$ has
inclination angle $i_{\rm e}$ with the line of sight (see Fig~\ref{fig1}).
An azimuthal angle $\phi_{\rm env}$ describes the orientation of the
projected $OZ'$ axis in the $x-y$ plane, which is the plane of the sky
for the observer.  The density prescription for the envelope becomes

\begin{equation}
n_{l'{\rm m'}}(r) = 2\pi\,(R_1-R_2)\,n_0\,K_{l'}\,Y_{l'{\rm m'}}
    (i_{\rm e},\phi_{\rm e}),
    \label{eq:nellipse}
\end{equation}

\noindent where

\begin{equation}
n(r,\vartheta',\varphi') = \left\{ \begin{array}{cl} n_0 & \hspace{3ex}\mbox{\rm when}
    \; r_2(\mu)\le r\le r_1(\mu), \\ 0 & \hspace{3ex}\mbox{\rm otherwise.}
    \end{array}\right. 
\end{equation}

\noindent Here $n_0$ is the {\em uniform} number density of particles within
the column bounded by $r_1$ and $r_2$, where

\begin{equation}
r_{1,2} (\mu) = \frac{R_{1,2}}{\sqrt{1+(A_{\rm r}^2-1)\mu^2}},
\end{equation}

\noindent and

\begin{equation}
\mu=\cos{\zeta}.
\end{equation}

\noindent The lengths $R_1$ and $R_2$ are the outer and inner {\em
equatorial} axis lengths, and $A_{\rm r}$ is the ratio of the length
of the equatorial width to the polar width (see Fig.~\ref{fig1}).
The angle $\zeta$ is the angle between the radius vector and the axis
of symmetry, which is related to our frames by the addition theorem
of spherical harmonics.  This explains the appearance of $Y_{l'{\rm
m'}}(i_{\rm e},\phi_{\rm e})$ in Eq.~(\ref{eq:nellipse}) -- see Simmons
(1982) and Jackson (1975).  Finally, $K_{l'}$ is given by

\begin{equation}
K_{l'} = \int_{-1}^1 \, \frac{P_{l'} (\mu)}{\sqrt{1+(A_{\rm r}^2-1)\mu^2}}
    \,d\mu, \label{eq:kl}
\end{equation}

\noindent where $P_{l'}(x)$ are Legendre polynomials.

Inserting Eqs.~(\ref{eq:Fellipse})--(\ref{eq:kl}) into
Eq.~(\ref{eq:Sfunc}), we obtain

\begin{equation}
S_{ll'{\rm mm'}} = I_*\,N\,f_{l{\rm m}}\,K_{l'}\,Y_{ll'{\rm m'}}
    (i_{\rm e},\phi_{\rm e}),
\end{equation}

\noindent where a convenient column density $N$ is introduced as

\begin{equation}
N= 2\pi\,(R_1-R_2)\,n_0,
\end{equation}

\noindent and

\begin{equation}
f_{l{\rm m}} = \int_0^\pi\,\int_0^{2\pi}\,A_{\rm p}(\vartheta,\varphi)\,
    Y^*_{l{\rm m}}(\vartheta,\varphi)\,\sin\vartheta\,d\vartheta\,
    d\varphi.
\end{equation}

\noindent Note that the multipoles of the flux $f_{l{\rm m}}$ are now
functions of the star's effective shape.  The multipole coefficients of
the envelope $K_{l'}$ are functions of the envelope's shape and size.
Conceptually, the product of $K_{l'}$ and $f_{l{\rm m}}$ leads to an
overall net amplitude ``pattern'' of scattered light as determined by
whether the two functions enhance or offset one another.  This pattern
sets the contribution level to the polarization and scattered flux as a
function of direction about the source.  The scattering phase function
(in our case dipole scattering) along with appropriate rotations operates
as a weighted filter function that acts on the amplitude pattern,
converting it to a net observed Stokes vector signal from the source
and envelope system.

Describing the orientation of the star relative to the observer frame
in terms of the Euler angles, we can choose $\alpha$ as zero, $\beta$
as the viewing inclination ($i_s$) of the $OZ-$axis (i.e., the nominal
rotation axis of the star) to the line of sight $Oz$, and $\gamma$
as the azimuth of the $OZ-$axis from the $Ox-$axis as measured about
the $Oz-$axis, which we denote as $\phi_{\rm s}$.  Thus $\phi_{\rm s}$
measures the rotational phase of the star relative to the observer
(the same conventions as adopted in Paper~I; see their Fig.~1).

In the observational context, it is the normalized scattered flux and
Stokes parameters that are usually used.  These are given by $(f_{\rm sc}, q,
u) = (F_{\rm sc}, Q, U)/ F_{\rm tot}$, where the total flux received
$F_{\rm tot}$ comprises the combination of the scattered flux $F_{\rm
sc}$ plus direct flux from the star $F_\ast$.  This contribution by direct
stellar light we denote as $F_\ast = I_*\,A_{\rm p}(i_{\rm s},\phi_{\rm
s})/D^2$.  So the total observed flux becomes

\begin{equation}
F_{\rm tot} = \frac{I_\ast}{D^2}\,\left[ A_{\rm p}(i_{\rm s},\phi_{\rm s})
    +(D^2/I_\ast)\,F_{\rm sc} \right] .
\end{equation}

\noindent The general expressions for the normalized scattered flux
and stokes parameters become

\begin{eqnarray}
f_{\rm sc} & = & \frac{\tau}{4\pi\,D^2\,F_{\rm tot}}\,\sum_{l {\rm m n}} \, R^l_{\rm nm}
        (0,i_{\rm s},\phi_{\rm s}) \, f_{l{\rm m}}\times \nonumber \\
 & & \left[ \sqrt{4\pi}C_{ll'{\rm nm'}}^{00} + \sqrt{\frac{\pi}{5}}
        C_{ll'{\rm nm'}}^{20}\right]
        \label{eq:fnorm}
\end{eqnarray}

\noindent and

\begin{eqnarray}
z^* & = & \frac{3\tau}{4\pi\,D^2\,F_{\rm tot}}\,\sqrt{\frac{2\pi}{15}}\sum_{l {\rm m n}}
        \, R^l_{\rm nm}(0,i_{\rm s},\phi_{\rm s}) \, f_{l{\rm m}} \nonumber \\
 & & \sum_{l' {\rm m'}}K_{l'}\,Y_{l'{\rm m}}(i_{\rm e},\phi_{\rm e})\,
    C_{ll'nm'}^{22}\,S_{ll'mm'}, \label{eq:znorm}
\end{eqnarray}

\noindent where $\tau = \sigma (R_1-R_2)n_0$ is the equatorial optical
depth.  The degree of polarization is given by $p=|z^*| = |z|$, and the
polarization position angle is given by $\psi = \frac{1}{2} arg \,z$.

From Eqs.~(\ref{eq:fnorm}) and (\ref{eq:znorm}), we can calculate
the polarization and the scattered flux, using the properties of the
two factors $K_{l'}$ and $f_{l{\rm m}}$ that describe the envelope
geometry and source anisotropy.  Due to the symmetry of the 
functions chosen to describe the stellar flux and the scatterer
density distribution, $K_{l'}$ and $f_{l{\rm m}}$ are non-zero only
for even $l'$ and $l$, respectively.  

\section{Applications}


As mentioned above, the multipoles of the flux $f_{l{\rm m}}$ are
non-zero only for even $l$, and the spherical harmonics for $l\ge
4$ are important mainly for fairly large angular distortions of the
star from sphericity.  As
an example, Figures~\ref{fig2}a and b show the variation of $f_{l0}$
for an oblate/prolate star distorted only along its $c-$axis.  

For $l=l'=2$, the scattered flux is a sum of terms given by

\begin{eqnarray}
F_{\rm sc} & = & \frac{\sigma}{4\pi\,D^2}\,\left\{ \sqrt{4\pi}\,\left[
    \sum_{\rm m=-2}^{\rm m=2}\,S_{\rm m}^{22\,2-2}\, C^{00}_{22\,2-2}
    \right. \right. \nonumber \\
 & & + \sum_{\rm m=-2}^{\rm m=2}\,S_{\rm m}^{22\,1-1}\, C^{00}_{22\,1-1}
    \sum_{\rm m=-2}^{\rm m=2}\,S_{\rm m}^{22\,00}\, C^{00}_{22\,00}
    \nonumber \\
 & & + \sum_{\rm m=-2}^{\rm m=2}\,S_{\rm m}^{22\,-11}\, C^{00}_{22\,-11}
        \sum_{\rm m=-2}^{\rm m=2}\,S_{\rm m}^{22\,-22}\, C^{00}_{22\,-22}
        \nonumber \\
 & & + S_{\rm m}^{00\,00}\, C^{00}_{00\,00} \left] + \sqrt{\frac{\pi}{5}}
    \, \right[ S_0^{02\,00}\, C^{20}_{02\,00} \nonumber \\
 & & + \sum_{\rm m=-2}^{\rm m=2}\,S_{\rm m}^{22\,2-2}\, C^{20}_{22\,2-2}
        \sum_{\rm m=-2}^{\rm m=2}\,S_{\rm m}^{22\,1-1}\, C^{20}_{22\,1-1}
        \nonumber \\
 & & + \sum_{\rm m=-2}^{\rm m=2}\,S_{\rm m}^{22\,00}\, C^{20}_{22\,00}
        \sum_{\rm m=-2}^{\rm m=2}\,S_{\rm m}^{22\,-11}\, C^{20}_{22\,-11}
        \nonumber \\
 & & \left. \left. + \sum_{\rm m=-2}^{\rm m=2}\,S_{\rm m}^{22\,-22}\, C^{20}_{22\,-22}
        \sum_{\rm m=-2}^{\rm m=2}\,S_{\rm m}^{20\,00}\, C^{20}_{20\,00}
    \right] \right\},   \label{eq:fresult}
\end{eqnarray}

\noindent and the Stokes parameters as

\begin{eqnarray}
Z^* & = &  \frac{3\sigma}{4\pi\,D^2}\,\sqrt{\frac{2\pi}{15}}\,\left\{
    S_0^{02\,02}\,C^{22}_{02\,02} \right. \nonumber \\
 & & + \sum_{\rm m=-2}^{\rm m=2}\,S_{\rm m}^{22\,20}\,C^{22}_{22\,20}
    + \sum_{\rm m=-2}^{\rm m=2}\,S_{\rm m}^{22\,11}\,C^{22}_{22\,11} \nonumber \\
 & & \left. + \sum_{\rm m=-2}^{\rm m=2}\,S_{\rm m}^{22\,02}\,C^{22}_{22\,02}
        + \sum_{\rm m=-2}^{\rm m=2}\,S_{\rm m}^{20\,20}\,C^{22}_{20\,20} \right\},
    \label{eq:zresult}
\end{eqnarray}

\noindent where

\begin{equation}
S_{\rm m}^{ll'\,{\rm nm'}} = R^l_{\rm nm} (\alpha,\beta,\gamma)\,S_{ll'\,
    {\rm mm'}}.
\end{equation}

\noindent The various coefficients $C^{LM}_{ll' nm'}$ are tabulated
in the Appendix in Tables \ref{tab:t1}-\ref{tab:t3}.

The multipoles $f_{l{\rm m}}$ and $K_{l'}$ are the determining factors
for properties of the total scattered light. In Paper~I for spherical
envelopes, we focused on the effect of the projected area $A_{\rm p}$
(within the shape factor $f_{l{\rm m}}$), finding that the polarization
increases as the stellar inclination $i_{\rm s}$ increases toward more
edge-on viewing perspectives of the star.  With an ellipsoidal envelope,
the ratio of the length of the equatorial axis to the polar axis of the
ellipsoidal envelope $A_{\rm r}$ (contained within the shape factor
$K_{l'}$) adds additional richness to the problem.  For $A_{\rm r}$
approaching infinity, the envelope will be a planar disk of scattering
particles.  As $A_{\rm r}$ approaches zero, the envelope stretches to
a cylindrical column.  Note that such a form is not the same as a polar
jet, since in our envelope parametrization the equatorial radius of the
envelope is never zero.

It turns out that $K_0$ and $K_2$ are straightforwardly derivable
as functions of $A_{\rm r}$.  There are two sets of solutions, one
for prolate envelopes and the other for oblate ones.  For prolate
envelopes (i.e., $A_{\rm r}<1$), the solution is

\begin{eqnarray}
K_0 & = & 2\frac{\sin^{-1} \sqrt{1-A_{\rm r}^2}}{\sqrt{1-A_{\rm r}^2}} 
	\label{eq:proK0} \\
K_2 & = & \frac{3}{2}\,\frac{K_0-2A_{\rm r}}{1-A_{\rm r}^2} - K_0
	\label{eq:proK2} 
\end{eqnarray}

\noindent whereas for oblate envelopes (i.e., $A_{\rm r}>1$), the solution is

\begin{eqnarray}
K_0 & = & \frac{1}{\sqrt{A_{\rm r}^2-1}}\,\ln\left(\frac{A_{\rm r}
    +\sqrt{A_{\rm r}^2-1}}{A_{\rm r}
    -\sqrt{A_{\rm r}^2-1}}\right) 
	\label{eq:obK0} \\
K_2 & = & \frac{3A_{\rm r}}{2(A_{\rm r}^2-1)}-\frac{1+2A_{\rm r}^2}
    {4(A_{\rm r}^2-1)}\,K_0
	\label{eq:obK2} 
\end{eqnarray}

\noindent The $K_0$ and $K_2$ functions have interesting limiting
behavior.  For both prolate and oblate envelopes, the coefficients
achieve their limiting values as $A_{\rm r}$ approaches
unity, with $K_0 = 2$ and $K_2 = 0$.  The opposite limits are extreme
distortions.  As $A_{\rm r}$ approaches zero, a prolate envelope
stretches out to an infinitely long column with $K_0 = \pi/2$ and
$K_2 =\pi/4$.  Note implicitly that there is an infinite amount of
scattering mass, but it is infinitely far away, hence $K_0$ and
$K_2$ remain well-behaved in this limit.  As $A_{\rm r}$ becomes
large, an obate envelope becomes a thin sheet in the equatorial
plane.  The coefficients both tend toward $K_0 = K_2 = 0$.

We next consider applications of Eqs.~(\ref{eq:fresult}) and
(\ref{eq:zresult}) to two particular cases.  The first case deals with a
scenario initially predicted by Owocki, Cranmer, \& Gayley (1996) for the
mass-loss from a rapidly rotating hot star, and explored in more detail
for the shaping of nebula surrounding Luminous Blue Variables (LBVs)
by Dwarkadas \& Owocki (2002). For line-driving of the wind coupled with
the effects of gravity darkening, Dwarkadas \& Owocki (2002) predict that
the equatorial wind will be less massive and slower than the polar flow.
The second case deals with a a Roche lobe filling star for a binary system
surrounded by a scattering envelope. This has relevance to short period
binaries experiencing mass transfer, where one of the stars dominates
the luminous output for the waveband of consideration.

\begin{figure}[t]
\resizebox{\hsize}{!}{\includegraphics[width=17cm]{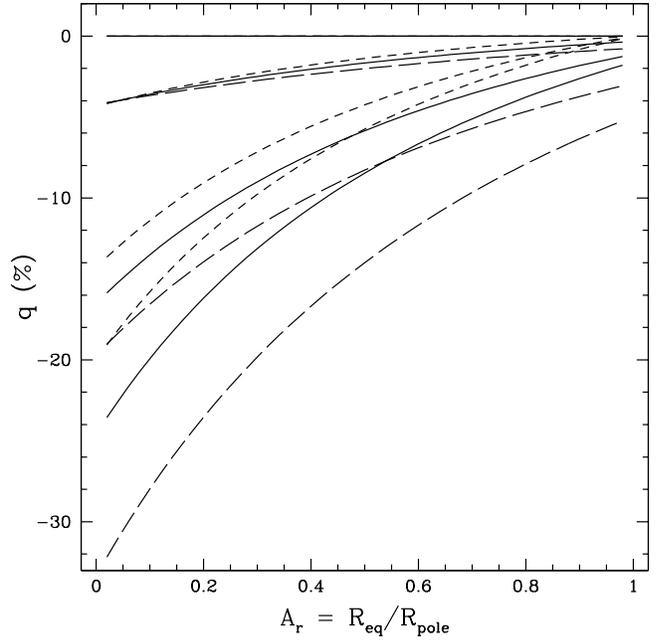}}
\caption{Polarization as a function of the prolate envelope shape
for different stellar anisotropies to represent the scenario of
an enhanced bipolar wind flow centered on a gravity darkened
star.  The curves are:  short dash for $c=1$; solid for $c=0.7$;
and long dash for $c=0.4$.  From bottom to top, the polarizations
are higher for more edge-on viewing inclinations (see the text).
The bipolar wind is aligned with the $c$ axis for the
source, hence $i_{\rm s}=i_{\rm e}$ in each case shown.  The
optical depth in the equator is fixed at $\tau_{\rm 0,eq}
= 2/\pi=0.64$.
\label{fig3}}
\end{figure}

\subsection{Rotating Stars}

For the case of a line-driven wind from a rotating star, we adopt the
following approximations to represent the polar enhanced density and
a reduced density from the equatorial band (c.f., Fig.~4 of Dwarkadas
\& Owocki 2002).  Although the wind density will fall off as $r^{-2}$
or faster, our main concern is to capture the trend of the latitude
dependence in the radial optical depth of scatterers, for which our
ellipsoidal envelope representation in the form of a prolate shape
is adequate.  With Thomson scattering being the dominant polarigenic
opacity, we introduce an electron scattering optical depth through the
equator of the envelope as $\tau_{0,{\rm eq}} = n_0\,(R_1-R_2)\,\sigma_T$.
Then the optical depth along any other radial line is

\begin{equation}
\tau(\mu) = \frac{\tau_{0,{\rm eq}}}{\sqrt{1-(1-A_{\rm r}^2)\mu^2}},
	\label{eq:taurot}
\end{equation}

\noindent where $A_{\rm r} <1$ leads to a higher optical depth of
scatterers along the pole than in the equator.  It is worth noting
that the optical depth averaged over solid angle is

\begin{equation}
\bar{\tau} = \tau_{0,{\rm eq}} \,\frac{\sin^{-1} \sqrt{1-A_{\rm r}^2}}
    {\sqrt{1-A_{\rm r}^2}}.
\end{equation}

\noindent The ratio factor ranges from unity for $A_{\rm r}=1$ (a
spherical envelope) to $\pi/2$ for $A_{\rm r}=0$ (a cylindrical
column).  Although in this latter case the optical depth along the
pole formally diverges, the average over solid angle $\bar{\tau}$
remains finite.

For a rapidly rotating star, gravity darkening leads to a lower
temperature at equatorial latitudes and polar brightening at high
latitudes.  With limb darkening included, a scattering particle at
an arbitrary position ``sees'' a complicated brightness pattern at
the star.  Above the pole the radiation field as seen by the scatterer
is centro-symmetric.  In the equatorial plane, the brightness pattern
of the star is banded, modulo the effect of limb darkening.  The
flux along the pole is therefore greater than in the equatorial
plane, and so we model this trend using equations~(\ref{eq:ptsource})
and (\ref{eq:areal}) with $a=b$ and $c<a$ as before, so the star
is oblate in terms of its illumination.

We apply our theory to this scenario by adopting $\tau_{0,{\rm eq}}=2/\pi$
to obtain rough upper limits to expected polarizations in the optically
thin approximation, since $\bar{\tau} \le 1$ for all $A_{\rm r} \le 1$.
Results are displayed in Figure~\ref{fig3} as a function of $A_{\rm
r}$ for different degrees of source anisotropy as characterized by
the value of $c$.  In this figure the polarization $q$ is plotted as
percent polarization and is normalized to the optical depth in the equator
$\tau_{\rm 0,eq}$.  The short dashed lines are for an isotropic source.
Solid lines are for $c=0.7$ and long dashed ones are for $c=0.4$.
Each case is shown at four viewing inclinations of $0^\circ, 30^\circ,
60^\circ$, $90^\circ$.  The larger polarizations $|q|$ are for higher
viewing inclinations.  With the envelope and star axes aligned, $i_{\rm s}
= i_{\rm e}$, thus the curves for the pole-on case have $q=0$
for all values of $A_{\rm r}$.

With enhanced mass loss and stellar illumination along the poles,
our model naturally predicts a polarization position angle that
is {\em orthogonal} to the rotation axis of the star.  Several LBV
stars have been studied and found to have (a) net polarizations and
(b) axisymmetric nebulae.  Schulte-Ladbeck \etal\ (1994) reported
on a polarimetric study of the LBV AG~Car, and they found strong
polarimetric variability.  Yet, the variable polarization displayed
a preferred axis co-aligned with the symmetry axis of the star's ring
nebula, exactly opposite of what our model would predict for enhanced
mass loss from the poles.  A more recent study by Davies \etal\ (2005)
arrived at a different conclusion.  Those authors found largely erratic
changes in polarization, in degree and position angle, similar to what is
observed in P~Cygni (Taylor \etal\ 1991). In fact, Davies \etal\ found
that three LBV stars had essentially random polarimetric variations
suggesting that these winds have no preferred symmetry axis but that
wind clumping could explain the variability.  On the other hand, their
discussion of $\eta$~Car is consistent with an enhanced bipolar wind,
for which the polarization position angle is indeed perpendicular to
the axis of the homunculus.  They also found that R127 shows evidence
for a symmetry axis.  Schulte-Ladbeck \etal\ (1993) concluded the same;
however, Davies \etal\ interpret the net polarization as evidence for
an enhanced {\em equatorial} density, not a bipolar enhancement.

Note that although our form for $\tau(\mu)$ in equation~(\ref{eq:taurot}) 
gives the correct trend with latitude about the star, the curve does not
in general accurately reproduce the shape expected of the Owocki \etal\
(1996) model.  In that paper the mass-loss rate and wind terminal speed
scale with the latitude-dependent effective gravity, which decreases
from the pole to the equator.  For rotation speeds that are not too close
to break-up, $A_{\rm r} \lesssim 1$, and the shape of $\tau(\mu)$ is a
relatively good match.  But for faster rotations, corresponding to lower
values of $A_{\rm r}$, agreement between equation~(\ref{eq:taurot}) and
an accurate treatment of the Owocki \etal\ bipolar winds is not good:
the effective opening angle for the bipolar wind in our model tends
to be too small.  As a result, our treatment tends to overestimate the
expected polarization, but it captures the overall qualitative trends.

It is clear that the polarimetric behavior of LBV stars is complex.
Some show evidence for axisymmetry, as one would expect from the
Owocki \etal\ (1996) mechanisms, and others do not.  They are certainly
strongly variable in their polarization, likely a consequence of wind
inhomogeniety.  The lack of a symmetry axis for some sources may simply be
the result of a range of rotation rates in the stars.  It is also likely
that the optically thin electron scattering may not be applicable to the
high mass-loss rate winds of many LBVs.  Large optical depths tend toward
depolarization relative to the thin limit because of multiple scattering.
The net polarization will thus be dominated by optically thin regions
(e.g., Taylor \& Cassinelli 1992).  If an LBV has a bipolar flow that
is thick to electron scattering, it may be that the equatorial flow
of lower optical depth could determine the polarization, leading to
a position angle that is parallel to the star's rotation axis instead
of perpendicular to it.  More modeling is needed to determine whether
the mechanism of Owocki \etal\ (1996) is consistent with the known
polarizations of LBV winds.

\subsection{Binaries}

\begin{figure}
\resizebox{\hsize}{!}{\includegraphics[width=17cm]{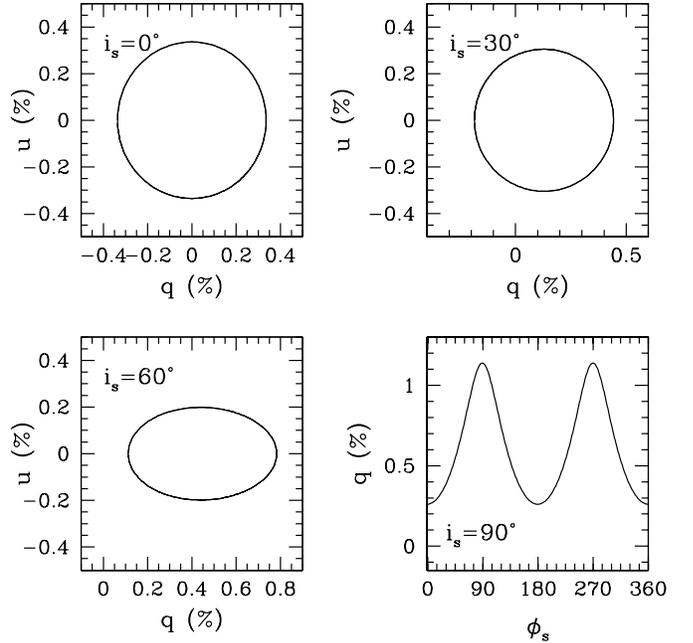}}
\caption{Example $qu$-diagrams for a prolate
illuminating source that rotates about an axis orthogonal to its
long axis. The polarizations are for an oblate circumstellar
envelope with symmetry axis coincident with the star's rotation
axis. Each panel is for a different viewing inclination as
indicated. For an edge-on view, $u=0$, so only the
$q$-polarization with rotational phase is shown. \label{fig4}}
\end{figure}

Binary systems provide a rich variety of scenarios involving distorted
stars and non-spherical circumstellar and/or circumbinary envelopes.
One example is $\beta$~Lyr in which the lower mass companion is a
late B~giant star and Roche lobe overflowing (Harmanec 2002).  In
this system the more massive star is embedded in a thick disk, and
there is even a jet whose origin appears offset from either star
(Harmanec \etal\ 1996; Hoffman, Nordsieck, \& Fox 1998).  The system
also sports a significant kilo-Gauss level magnetic field (Skulsky
1982; Leone \etal\ 2003).

In our exploration of variable polarization in binary stars, we adopt
fixed values of $b=2$ and $a=c=1$ to represent a highly distorted
prolate-shaped Roche-lobe filling star in the system.  Consider an
envelope with an axis of symmetry that is parallel to the orbital axis
of the binary (i.e., $i_{\rm e} = i_{\rm s}$).  Figure~\ref{fig4} shows
variable polarizations in the case of an oblate scattering envelope with
$A_{\rm r}=3$ and $\tau=0.1$.  The maximum polarization is about 1.1\%.
In Paper~I, this star model was considered with a spherically symmetric
scattering envelope, and the maximum polarization was only 0.45\%.
Clearly, both the anisotropy of the illuminating source and the distortion of
the scattering envelope from spherical are important for interpreting
polarimetric data from such systems.

\begin{figure}
\resizebox{\hsize}{!}{\includegraphics[width=17cm]{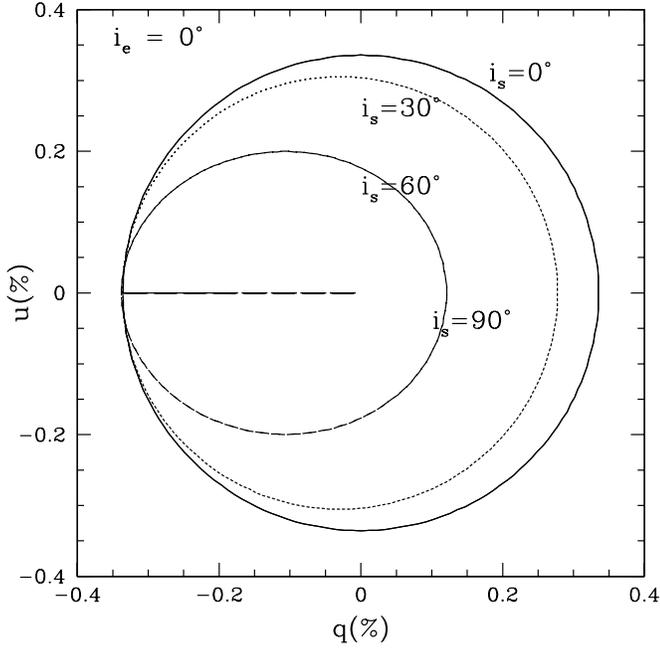}}
\caption{ The same prolate star as in
Fig.~\ref{fig4}, but now with the oblate envelope not co-axial with
the rotation axis. Here the envelope is held fixed at pole-on, and
the source is allowed to tilt as indicated. \label{fig5}}
\end{figure}

\begin{figure}
\resizebox{\hsize}{!}{\includegraphics[width=17cm]{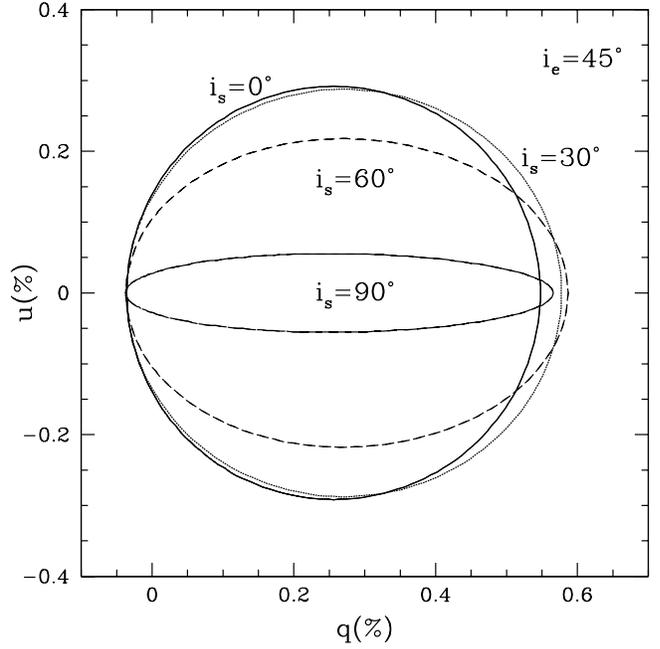}}
\caption{As in Fig.~\ref{fig6}, now with the envelope held fixed
at an intermediate viewing inclination of $i_{\rm e} = 45^\circ$.
\label{fig6}}
\end{figure}

\begin{figure}
\resizebox{\hsize}{!}{\includegraphics[width=17cm]{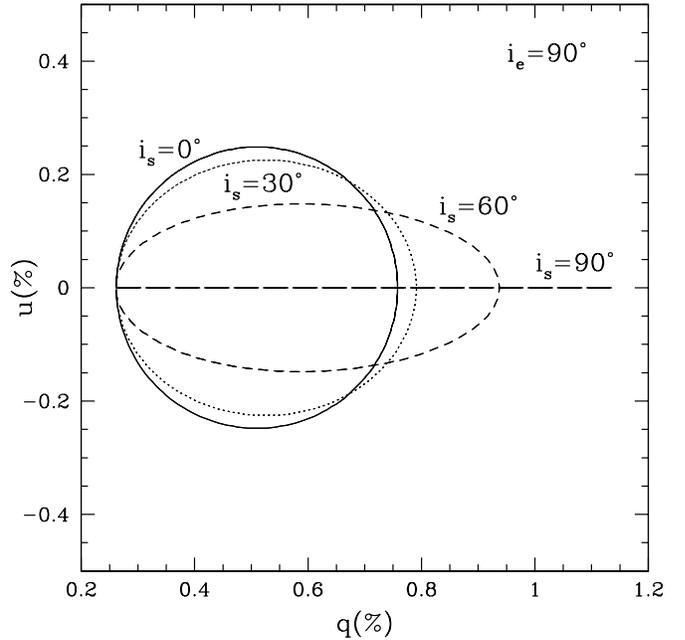}}
\caption{As in Fig.~\ref{fig6}, now with the envelope fixed
at an edge-on view.
\label{fig7}}
\end{figure}

In some cases there is evidence that the axis of reference for
the scattering envelope is not aligned with the reference axis for
the star.  This can occur in stars with strong magnetic fields,
such as oblique magnetic rotators (Stibbs 1950; Deutsch 1956), 
which are sometimes found
in close binary systems as well.  For a strong oblique
magnetic field, the envelope's reference axis will be parallel to
the field axis instead of the rotation axis (in this case for the
binary orbital motion).  Figures \ref{fig5}--\ref{fig7} shows
calculations for three envelope inclinations of $i_{\rm e} = 0^\circ$,
$45^\circ$, and $90^\circ$ respectively.  As in Figure~\ref{fig4},
the stellar illumination is still represented by the case $a=c=1$ and
$b=2$ The figures show $q$-$u$ diagrams at stellar inclinations of
$i_{\rm s} = 0^\circ$, $30^\circ$, $60^\circ$, and $90^\circ$. The
overall trend is for the inclination of the star to affect the
degree of polarization more substantially than the inclination of
the envelope. However, the largest polarization values generally
occur for the largest differences in the respective inclinations
$i_{\rm s}-i_{\rm e}$, because the stellar flux is greatest along
its smallest axes ($a$ and $c$), while the envelope possesses a greater
column density of 
scatterers along its longest axes.

%

\section{Conclusions}

In this work we derive the polarization arising from an
anisotropic point light source within an arbitrary shaped envelope. 
We specifically considered Thomson and Rayleigh
scattering mechanisms. The mathematical analysis made use of
the properties of spherical harmonics, which can readily be
generalized to more complicated cases than those discussed here
(e.g., spots on a star). By using the
first few spherical harmonics, an approximation was derived to
represent distortions of stars and envelopes from spherical by varying
amounts. The anisotropy properties of the star and envelope can be
reduced to representations as products of multipole contributions.

We considered applications to rotating stars and binary
systems.  For rotating stars we sought to explore the polarizations
resulting from an approximate representation of enhanced bipolar
wind flow illuminated by a rapidly rotating and gravity darkened
star. This scenario has relevance to some LBV winds.

We also considered a highly distorted star such as can occur in a
binary when one of the components is Roche lobe filling and
dominates the luminous output of the system.  In this case the
star's illuminating characteristics were modeled as a prolate
ellipsoid rotating about an axis
perpendicular to its symmetry axis. Scenarios involving
an oblate circumstellar envelope that was aligned or inclined with
respect to the star's rotation axis were considered.

The theory presented here is fairly general; however, there remain a
number of improvements that should be pursued.  Foremost is
inclusion of the star's finite size, because of both occultation and
the finite depolarization effects.  An initial attempt at this has
been made by Ignace \etal\ (2008), who consider 
scattering polarization from a structured giant star chromosphere.
In addition, our theory assumes that the radial and angular
descriptions of the envelope are separable, which may not always be the
case.  Including these modifications 
is far from trivial but may be necessary in order to accurately
represent complex systems.  Certainly, the growth in the size and
capability of telescopes and modern instrumentation, along with the
increased importance of spectropolarimetry in ascertaining source
geometries (e.g., supernova ejecta -- Wang \etal\ 2001; Leonard
\etal\ 2006) motivates an effort to generalize further our approach.

\appendix

\section*{Appendix}

The Clebsh-Gordon coefficients $C^{\rm LM}_{ll'\,{\rm mm'}}$ arise from
the products of two spherical harmonics and are given by (c.f.,
Messiah 1962)

\begin{eqnarray}
C^{\rm LM}_{ll'\,{\rm nm'}} & = & (-1)^{\rm M}\,\sqrt{\frac{(2l+1)(2l'+1)
        (2L+1)}{4\pi}} \times \nonumber\\
 & & \left( \begin{array}{ccc} l & l' & {\rm L} \\ 0 & 0
        & 0 \end{array} \right)\,\left( \begin{array}{ccc} l & l' & {\rm L} \\
        {\rm n} & {\rm m'} & {\rm M}\end{array} \right)
        \label{eq:Aclebsh}
\end{eqnarray}

\noindent where

\begin{eqnarray}
\left( \begin{array}{ccc} a & b & c \\ \alpha & \beta & \gamma \end{array}
    \right) & = &( -1 )^{a-b-\gamma}\,\sqrt{\Delta(abc)}\nonumber \\
  & \times &\sqrt{ (a+\alpha)!\,(a-\alpha)!\,(b+\beta)!\,(b-\beta)!\,
    (c+\gamma)!\,(c-\gamma)! } \nonumber \\
  & \times &\left\{ \sum_t \,(-1)^t\,\left[ \, t!\,(c-b+t+\alpha)!\,
    (c-b+t-\beta)! \right. \right. \nonumber \\
  & \times &\left. \left. (a+b-c-t)!\,(a-t-\alpha)! (b-t-\beta)!
    \, \right] \right\}.	\label{eq:racah}
\end{eqnarray}

\noindent with

\begin{equation}
\Delta(abc) = \frac{(a+b-c)!\,(b+c-a)!\,(c+a-b)!}{(a+b+c+1)!},
\end{equation}

\noindent and $t$ is an integer value for which the arguments of the
factorials are positive or zero.  The number of terms in this sum
is $1+\eta$, where $\eta$ is the smallest of the nine numbers $a\pm
\alpha$, $b\pm \beta$, $c\pm\gamma$, $a+b-c$, $b+c-a$, and $c+a-b$
(c.f., Messiah 1961).

Expression~(\ref{eq:racah}) is called the Racah formula and is non-zero
under the following two conditions:

\begin{equation}
\alpha + \beta + \gamma = 0,
\end{equation}

\noindent and

\begin{equation}
|a-b| \le c \le a+b.
\end{equation}

\noindent Values of the Clebsch-Gordon coefficients used in this paper
are given in Tables \ref{tab:t1}--\ref{tab:t3}.

\begin{table}
\caption{Values of $C^{00}_{ll'{\rm mm'}}$  \label{tab:t1} }
\begin{tabular}{|c||c|c|c|}
\hline  mm'/\ $ll'$ & 2,2 & 1,1 & 0,0 \\ \hline \hline
2,$-$2 & $1/\sqrt{4\pi}$ & 0  & 0  \\ \hline
1,$-$1 & $-1/\sqrt{4\pi}$ &  $-1/\sqrt{4\pi}$ &    \\ \hline
0,0 & $1/\sqrt{4\pi}$ &  $1/\sqrt{4\pi}$ &  $1/\sqrt{4\pi}$\\  \hline
$-$1, 1 & $-1/\sqrt{4\pi}$ &  $-1/\sqrt{4\pi}$ &   \\  \hline
$-$2, 2 & $1/\sqrt{4\pi}$ & 0  &  0\\  \hline
\end{tabular}
\end{table}

\begin{table}
\caption{Values of $C^{20}_{ll'{\rm mm'}}$  \label{tab:t2} }
\begin{tabular}{|c||c|c|c|c|c|c|}
\hline  mm'/\ $ll'$ & 2,0 & 2, 1 & 2,2 & 1,1 & 1,2 & 0,2 \\ \hline \hline
2,$-$2  & 0 & 0 & $-\frac{1}{\sqrt{4\pi}}\,\frac{12\sqrt{5}}{7}$ & 0 & 0  & 0  \\ \hline
1,$-$1  & 0 & 0 & $-\frac{1}{\sqrt{4\pi}}\,\frac{12\sqrt{5}}{7}$ &  $\frac{1}{\sqrt{4\pi}}\,\frac{4}{\sqrt{5}}$ &  0 & 0  \\ \hline
0,0   & $\frac{1}{\sqrt{4\pi}}$ & 0 & $\frac{1}{\sqrt{4\pi}}\,\frac{12\sqrt{5}}{7}$ &  $\frac{1}{\sqrt{4\pi}}\,\frac{2}{\sqrt{5}}$ &  0 & $\frac{1}{\sqrt{4\pi}}$ \\ \hline
$-$1, 1 & 0 & 0 & $-\frac{1}{\sqrt{4\pi}}\,\frac{12\sqrt{5}}{7}$ &  $\frac{1}{\sqrt{4\pi}}\,\frac{4}{\sqrt{5}}$ &  0 & 0  \\ \hline
$-$2, 2 & 0 & 0 & $-\frac{1}{\sqrt{4\pi}}\,\frac{12\sqrt{5}}{7}$ & 0  & 0 &  0 \\ \hline
\end{tabular}
\end{table}

\begin{table}
\caption{Values of $C^{22}_{ll'{\rm mm'}}$  \label{tab:t3} }
\begin{tabular}{|c||c|c|c|c|c|c|}
\hline  mm'/\ $ll'$ & 2,0 & 2,1 & 2,2 & 1,1 & 1,2 & 0,2 \\ \hline \hline
2,0  & $\frac{1}{\sqrt{4\pi}}$ & 0 & $-\frac{1}{\sqrt{4\pi}}\,\frac{\sqrt{20}}{7}$ & 0 & 0  & 0  \\  \hline
1,1  & 0 & 0 & $\frac{1}{\sqrt{4\pi}}\,\frac{\sqrt{30}}{7}$ &  $\frac{1}{\sqrt{4\pi}}$ &  0 & 0  \\  \hline
0,2   & 0 & 0 & $-\frac{1}{\sqrt{4\pi}}\,\frac{\sqrt{20}}{7}$ &  0 &  0 & $\frac{1}{\sqrt{4\pi}}$ \\ \hline
\end{tabular}
\end{table}

\begin{acknowledgements}

RI is grateful for funding from the National Science Foundation, grant
AST-0807664.  This work has been supported financially by the University
of King Saud and by a grant from the Science and Engineering Research
Council.  In addition, JCC acknowledges funding from a partnership
between the National Science Foundation (NSF AST-0552798), Research
Experiences for Undergraduates (REU), and the Department of Defense
(DoD) ASSURE (Awards to Stimulate and Support Undergraduate Research
Experiences) programs. Finally, we also thank Gary Henson for his many
helpful discussions.

\end{acknowledgements}



\clearpage

\begin{thebibliography}{}

\bibitem {} Al-Malki, M.~B., Simmons, J.~F.~L., Ignace, R., Brown, J.~C.,
    Clarke, D., 1999, A\&A, 347, 919
\bibitem {} Bjorkman, J.~E., Bjorkman, K.~S., 1994, ApJ, 436, 818
\bibitem {} Brown, J.~C., Carlaw, V.~A., Cassinelli, J.~P., 1989,
    ApJ 344, 341
\bibitem {} Brown, J.~C., Fox, G.~K., 1989, ApJ 347, 468
\bibitem {} Brown, J.~C., Ignace, R., Cassinelli, J.~P., 2000, A\&A, 
	356, 619
\bibitem {} Brown, J.~C., McLean, I.~S., 1977, A\&A 57, 141
\bibitem {} Brown, J.~C., McLean, I.~S., Emslie, A.~G., 1978
    A\&A 68, 415
\bibitem {} Cassinelli, J.~P., Nordsieck, K.~H., Murison, M.~A., 1987,
    ApJ 317, 293
\bibitem {} Clarke, D., McGale, P.~A., 1986, A\&A 169, 251
\bibitem {} Clarke, D., McGale, P.~A., 1987, A\&A 178, 294
\bibitem {} Collins, G.~W., II, 1970, ApJ, 159, 583
\bibitem {} Davies, B., Vink, J.~S., Oudmaijer, R.~D., 2007, A\&A, 469, 1045
\bibitem {} Davies, B., Oudmaijer, R.~D., Vink, J.~S., 2005, A\&A, 439, 1107
\bibitem {} Deutsch, A.~J., 1956, PASP, 68, 92
\bibitem {} Dwarkadas, V.~V., Owocki, S.~P., 2002, ApJ, 581, 1337
\bibitem {} Dyck, H.~M., Frobes, F.~F., Shawl, S.~J., 1971, ApJ 76, 901
\bibitem {} Elias, N.~M., Koch, R.~H., Pfeiffer, R.~J., 2008, A\&A, 489, 911
\bibitem {} Friend, D.~F., Cassinelli, J.~P., 1986, ApJ 306, 215
\bibitem {} Fox, G.~K., 1991, ApJ 379, 663
\bibitem {} Fox, G.~K., Brown, J.~C., 1991, ApJ 375, 300
\bibitem {} Hamann, W.-R., Feldmeier, A., Oskinova, L.~M., 2008, Clumping
in Hot-Star Winds, (Universit\"{a}tsverlag Potsdam)
\bibitem {} Harmanec, P., 2002,
    Astronomische Nachrichten, 323, 87
\bibitem {} Harmanec, P., Morand, F., Bonneau, D., Jiang, Y., Yang, S.,
    \etal, 1996, A\&A, 312, 879
\bibitem {} Hoffman, J.~L., Nordsieck, K.~H., Fox, G.~K., 1998, AJ, 115, 1576
\bibitem {} Ignace, R., Henson, G.~D., Carson, J., 2008, to appear in
    The Biggest, Baddest, Coolest Stars, (eds) D.~Luttermoser, B.~J.\
    Smith, R.~Stencel, (ASP Conf.\ Ser.)
\bibitem {} Jackson, J.~D., 1975, Classical Electrodynamics, 2nd ed.,
    J.~Wiley \& Son, New York
\bibitem {} Kruszewski, A., 1968, PASP, 80, 560
\bibitem {} Leone, F., Plachinda, S.~I., Umana, G., Trigilio, C.,
    Skulsky, M., 2003, A\&A, 405, 223
\bibitem {} Li, Q., Brown, J.~C., Ignace, R., Cassinelli, J.~P., Oskinova,
    L.~M., 2000, A\&A, 357, 233
\bibitem {} Leonard, D.~C., \etal\ 2006, Nature, 440, 505
\bibitem {} Messiah, A., 1962, Quantum Mechanics VII, North-Holland
    Publishing Company, Amsterdam
\bibitem {} Owocki, S.~P., Cranmer, S.~R., Gayley, K.~G., 1996, ApJ, 472,
    L115
\bibitem {} Patel, M., Oudmaijer, R.~D., Vink, J.~S., Bjorkman, J.~E.,
	Davies, B., \etal, 2008, MNRAS, 385, 967
\bibitem {} Richardson, L.~L., Brown, J.~C., Simmons, J.~F.~L., 1996, A\&A,
    306, 519
\bibitem {} Rudy, R.~J., Kemp, J.~C., 1978, ApJ 221, 220
\bibitem {} Schulte-Ladbeck, R., R.~E., Clayton, G.~C., Hillier, D.~J.,
    Harries, T.~J., Howarth, I.~D., 1994, ApJ, 429, 846
\bibitem {} Schulte-Ladbeck, R., R.~E., Leitherer, C., Clayton, G.~C.,
    Robert, C., Meade, M.~R., Drissen, L., Nota, A., Schmutz, W.,
    1993, ApJ, 407, 723
\bibitem {} Serkowski, K., 1970, ApJ 160, 1107
\bibitem {} Shawl, S.~J., 1975, apJ 80, 595
\bibitem {} Shakhovskoi, N.~M., 1965, Sov.\ Astr., 8, 833
\bibitem {} Simmons, J.~F.~L., 1982, MNRAS 200, 91
\bibitem {} Simmons, J.~F.~L., 1983, MNRAS 205, 153
\bibitem {} Skulsky, M.~Yu., 1982, Pis'ma Astron.\ Zh., 8, 238
\bibitem {} Stibbs, D.~W.~N., 1950, MNRAS, 110, 395
\bibitem {} Taylor, M., Cassinelli, J.~P., 1992, Apj, 401, 311
\bibitem {} Taylor, M., Nordsieck, K.~H., Schulte-Ladbeck, R.~E., Bjorkman,
    K.~S., 1991, AJ, 102, 1197
\bibitem {} van de Hulst, H.~C., 1957, Light Scattering by Small Particles,
    J.~Wiley \& Son, New York
\bibitem {} Vink, J.~S., Drew, J.~E., Harries, T.~J., Oudmaijer, R.~D.,
	2005, in Astronomical Polarimetry: Current Status and Future
	Directions, (eds.) A.\ Adamson, C.\ Aspin, C.~J.\ Davis, and T.\
	Fujiyoshi, ASP Conf.\ Ser.\ v.~343, 232
\bibitem {} Wang, L., Howell, D.~A., H\"{o}flich, P., Wheeler, J.~C.,
    2001, ApJ, 550, 1030

\end{thebibliography}
\end{document}